\documentclass[10pt,british,english,conference]{IEEEtran}
\usepackage[T1]{fontenc}
\usepackage[latin9]{inputenc}
\usepackage{float}
\usepackage{amsmath}
\usepackage{graphicx}
\usepackage{amssymb}

\makeatletter

\floatstyle{ruled}
\newfloat{algorithm}{tbp}{loa}
\floatname{algorithm}{Algorithm}

\ifCLASSINFOpdf
\else
\fi
\makeatother

\usepackage{babel}

\begin{document}

\title{Combinatorial Channel Signature Modulation for Wireless ad-hoc Networks}

\author{\IEEEauthorblockN{Robert J. Piechocki} \IEEEauthorblockA{Merchant
Venturers School of Engineering\\
 University of Bristol\\
 Woodland Rd, Bristol, BS8 1UB, UK\\
 Email: r.j.piechocki@bristol.ac.uk} \and\IEEEauthorblockN{Dino
Sejdinovic} \IEEEauthorblockA{Gatsby Computational Neuroscience
Unit\\
 University College London\\
 17 Queen Square, London, WC1N 3AR, UK\\
 Email: dino@gatsby.ucl.ac.uk} }
\maketitle
\begin{abstract}
In this paper we introduce a novel modulation and multiplexing method
which facilitates highly efficient and simultaneous communication
between multiple terminals in wireless ad-hoc networks. We term this
method Combinatorial Channel Signature Modulation (CCSM). The CCSM
method is particularly efficient in situations where communicating
nodes operate in highly time dispersive environments. This is all
achieved with a minimal MAC layer overhead, since all users are allowed
to transmit and receive at the same time/frequency (full simultaneous
duplex). The CCSM method has its roots in sparse modelling and the
receiver is based on compressive sampling techniques. Towards this
end, we develop a new low complexity algorithm termed Group Subspace
Pursuit. Our analysis suggests that CCSM at least doubles the throughput
when compared to the state-of-the art.
\end{abstract}

\section{Introduction}

Time dispersion has traditionally posed a very challenging problem
for communications systems. Typical examples of highly time dispersive
channels include wireless systems with large bandwidth, power line
communication (e.g. for Smart Grids), underwater channels etc. The
currently favoured state-of-the-art solution is typified by OFDM and
SC-FDE systems (e.g., 4G mobile systems, WiFi). Other existing solutions
include: equalisation in single carrier receivers (e.g., 2G mobile
systems) and rake receivers for CDMA (e.g., 3G mobile systems). In
all those techniques time dispersion represents a hindrance to a larger
or smaller extent. The system described here thrives on the dispersive
nature of communications channels and turns it into an advantage. 

MAC Layer coordination is another source of inefficiencies in communications
systems. The MAC protocol regulates how competing users access a shared
resource (e.g. a radio channel). In a standard solution only a single
user can occupy a shared resource; otherwise a \textquotedblleft{}collision\textquotedblright{}
occurs. The most important MAC protocols include CSMA/CA (e.g. IEEE
802.11x) or (slotted) Aloha. The DS-CDMA system somewhat relaxes this
constraint by allowing a group of synchronised users to transmit at
the same time and in the same frequency (in the same cell). However,
synchronisation is very difficult to achieve in an ad-hoc network.
The CCSM method does not require a complicated MAC layer coordination
mechanism. The CCSM allows all users to transmit signals at the same
time, therefore no coordination is needed. Another highly beneficial
feature is the ability to achieve a true duplex, i.e. all users in
the network can transmit and receive signals at the same frequency
and in the same time slot.

The CCSM method is inspired by a cross-layer scheme for wireless peer-to-peer
mutual broadcast considered by Zhang and Guo in \cite{peer-peer}.
In this paper each node is assigned a codebook of on-off signalling
codewords, such that every possible message corresponds to a single
codeword. However, the scheme by Zhang and Guo is suitable only {}``for
the situation where broadcast messages consist of a relatively small
number of bits''. Namely, the size of the sparse recovery problem
which needs to be solved is exponential in the length of the message.
Our scheme overcomes this limitation by encoding the message in a
combination of the codeword span, i.e., in a choice of $l$ out of
$L$ codewords in the codeword span, where $l\ll L$. Such representation
of useful information results in a significant reduction of the computational
complexity%
\footnote{In the set-up by Zhang and Guo, the size of the sparse vector to be
recovered is $L\cdot N$, where $L$ is the number of all possible
messages, and $N$ is the number od users. This means that each message
has $\log L$ nats of information. On the other hand, the same size
of the problem in our scheme results in the message length of $\log\binom{L}{l}$
nats of information for appropriately chosen $l\ll L$. If, for example,
$l=L^{1/2}$, the standard bounds on the binomial coefficients yield
$\log\binom{L}{l}=\mathcal{O}(L^{1/2}\log L^{1/2})$. Assuming $N$
fixed, the sparse recovery problem size is now only quadratic in the
number of nats of information per message. %
}, as the number of possible messages is expressed through a number
of all possible combinations, which is $\binom{L}{l}$. This, in turn,
renders our scheme practical for broadcasting much longer messages.
Moreover, in CCSM additional information can be encoded in the choice
of the weights assigned to a particular combination of the codeword
span. In addition, the scheme of \cite{peer-peer} cannot cope with
time dispersive environments. Our scheme, in contrast, thrives on
dispersive nature of wireless systems, by adapting the sparse recovery
problem to the channel signatures.

Combinatorial modulation constructions have been previously considered
in optical communication systems. A throughput efficient version of
pulse-position modulation (PPM) signalling scheme is called multipulse
or combinatorial PPM (MPPM) \cite{Sugiyama89,Budinger93,Georghiades94}.
However, MPPM applies such information representation directly in
time domain using single pulses. The MPPM signalling is inherently
sensitive to multipath interference, time dispersion and multiple
access interference (MAI) \cite{Zhao02}. Whereas MPPM signalling
typically uses a maximum-likelihood receiver \cite{Georghiades94},
which involves an optimisation problem over the set of all binary
sequences of length $L$ having weight $l$, which becomes intractable
even for moderate values of $L$ and $l$, the CCSM method utilizes
fast reconstruction methods based on sparse recovery solvers \cite{cosamp,SP}
found in the field of compressed sensing \cite{candes,donoho}.

\section{Signaling Modulation and Codebook Design}

\subsection{System Overview}

To improve the clarity of presentation we describe our system using
toy examples in baseband signaling. However, the system is equally
applicable to pass-band signaling, which, in fact, we use in the following
sections. 

\begin{figure}
\noindent \begin{centering}
\includegraphics[width=3.7in]{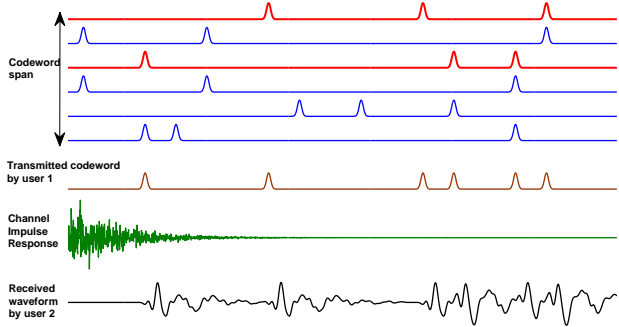}
\par\end{centering}

\caption{Simple example of a codebook and construction of the transmitted signal.\label{fig:Codebook}}

\end{figure}

Each of the users constructs its transmitted signal using a codeword
span known to all intended receivers. Figure \ref{fig:Codebook} depicts
an example of the codeword span with $L=6$. The message to be transmitted
is encoded in an $l$-\emph{combination} of the codeword span, i.e.,
in a choice of $l$ out of $L$ codewords in the codeword span, where
$l\ll L$. Note that there are $\left({L\atop l}\right)=\frac{L!}{l!\left(L-l\right)!}$
such combinations. Specifically, the transmitted signal is a weighted
sum of the chosen waveforms. In base-band, the weights could be points
in Amplitude Shift Keying (ASK) modulation e.g. $\left\{ +1,-1\right\} $.
In the provided example in Figure \ref{fig:Codebook}, $l=2$ waveforms
are chosen: first and third (depicted in red). Both weights happen
to be $+1$. The transmitted waveform is the sum of the two (brown
line). The information rate of this signaling scheme is thus $R=\frac{1}{W}\left(\log_{2}\left({L\atop l}\right)+lq\right)$
bits/s, where $W$ is the time duration of the waveforms in seconds,
and $2^{q}$ is the size of the alphabet of weights. 

This particular construction of constituent waveforms (codeword span),
combinatorial construction of the transmitted signal and the fact
that $l\ll L$ all play a crucial part since they allow very efficient
decoding, MAC-less user coordination and full duplex operation for
each user. A key feature of the constituent waveforms is sparsity
i.e. the waveforms are constructed from very short bursts of digital
modulation signals. We emphasize, it is not the digital signal which
carries useful information - the information rate is the same no matter
what modulation (BPSK, QPSK, 16-QAM etc) we choose to construct the
waveforms. It is the \emph{choice} of the $l$-combination of the
codeword span and of the associated weights which carries the information. 

The transmitted waveform is propagated in a dispersive channel (depicted
as a green line) and received as a convolution of the two (black line).
The implicit assumption here is that the channel can be modelled as
a linear time invariant channel (FIR filter). Such an assumption is
a commonplace in the literature and in practice. 

\begin{figure}
\noindent \begin{centering}
\includegraphics[width=3.7in]{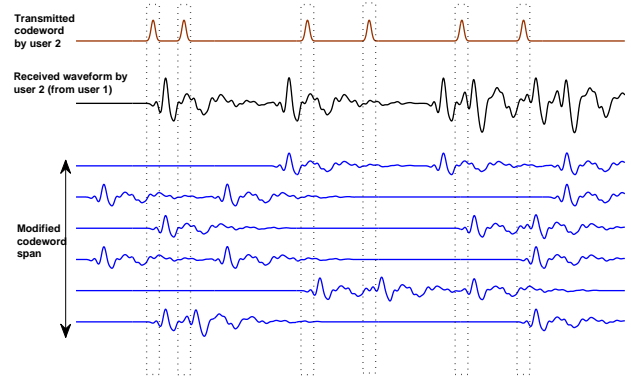}
\par\end{centering}

\caption{Simple example of a receiver codebook.\label{fig:Receiver-codebook}}

\end{figure}

The CCSM method relies on the linearity property of convolution. The
receiver reconstructs a modified codeword span \textendash{} blue
waveforms in Figure \ref{fig:Receiver-codebook}, where each waveform
in the original codeword span is convolved with the channel signature.
The task for the receiver is to estimate which $l$ waveforms were
used by the transmitter. The whole detection process can be performed
efficiently using sparse recovery solvers. The transmitted waveform
is essentially a sequence of on-off duty cycles, where for most of
the time there are silent periods ({}``off cycles''). Each user
utilises its {}``off cycles'' to receive signals from the other
users. In the \textquotedblleft{}on cycles\textquotedblright{}, however,
the user cannot receive the signal, which represents an erasure in
the codebook. This is depicted in Figure \ref{fig:Receiver-codebook}
as the doted boxes. Only non-erased portions of the codebook are used
in the detection process. Technically, with this scheme the Rx/Tx
chains do not operate simultaneously. Furthermore, the explicit assumption
is that the nodes operate fast switching (at the symbol rate) between
Rx/Tx, which is indeed possible with the current RF technology.

\subsection{Constant Weight Codes}

The CCSM requires a non-linear encoding operation. The process of
mapping the information vectors at each user to a unique $l$-combination
of the codeword span can be viewed as constant (hamming) weight coding
(CWC). The problem of efficient encoding and decoding constant-weight
vectors received significant interest in the literature. There are
practical algorithms of computational complexity linear in the length
$L$ of constant weight vectors, which are based on lexicographic
ordering and enumeration \cite{Ramabadran}. However, the approach
particularly suitable for our system is that of \cite{Tian_et_al},
as its complexity is quadratic in the weight of constant weight vectors,
which fares favourably in comparison to the enumeration approach in
the case where $l\ll L$. In \cite{Tian_et_al}, authors pursue geometric
representation of information vectors in an $l$-dimensional Euclidean
space and establish bijective maps by dissecting certain polytopes
in this space.

\subsection{CCSM Encoder}

Consider a network of $N+1$ users denoted $0,1,\ldots,N$, each of
which has a $k+lq$ bit message to transmit to all others through
a wireless medium using the same single carrier frequency. Denote
by $M$ the number of transmissions, and by $\omega_{i}\in\mathbb{F}_{2}^{k+lq}$
the message at user $i$. It is assumed that users are equipped with
an encoder, which constitutes of bijective maps $\phi_{\mathcal{C}}$
and $\phi_{w}$. The first map, $\phi_{\mathcal{C}}\,:\,\mathbb{F}_{2}^{k}\to\mathcal{C}$,
maps $k$-bit binary words into an $(L,l)$ constant weight binary
code $\mathcal{C}\subseteq\{c\in\mathbb{F}_{2}^{L}:\; w_{H}(c)=l\}$.
The second map $\phi_{w}\,:\,\mathbb{F}_{2}^{lq}\times\mathcal{C}\to\mathbb{C}^{L}$
assigns complex-numbered values to the non-zero entries in a constant
weight binary codeword from $\mathcal{C}$. For simplicity, we may
assume that $\mathcal{C}$ consists of all possible ${L \choose l}$
constant weight codewords, in which case we can take $k=\lfloor\log_{2}{L \choose l}\rfloor$.
Each user $i$ is assigned a signaling dictionary $\mathbf{S}_{i}=\left(\mathbf{s}_{i,1},\mathbf{s}_{i,2},\ldots,\mathbf{s}_{i,L}\right)$,
where each $\mathbf{s}_{i,j}\in\mathbb{C}^{M}$ is a sparse column
vector. (Columns of the matrix $\mathbf{S}_{i}$ can be thought of
as sampled waveforms constituting the codeword span in Fig. \ref{fig:Codebook}.)
Each user has perfect knowledge of all $N+1$ signaling dictionaries.
Furthermore, each user $i$ has a perfect knowledge of the channel
impulse responses $\mathbf{h}_{i,j}\in\mathbb{C}^{M}$ of the channel
between users $j$ and $i$, and of its own channel impulse response
$\mathbf{h}_{i,i}\in\mathbb{C}^{M}$, which we refer to as a {}``\emph{self-channel}''.
({}``Self-channel'' can be thought of as a \textquotedblleft{}radar
return\textquotedblright{}, and its role is explained in the description
of the CCSM decoder.) 

\begin{figure}
\noindent \begin{centering}
\includegraphics[bb=50bp 40bp 680bp 420bp,clip,width=3.5in]{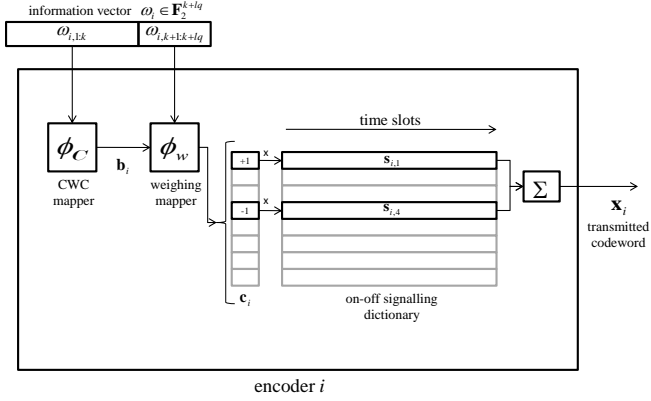}
\par\end{centering}

\caption{CCSM encoder at the $i$-th user.\label{fig:encoder-flow}}

\end{figure}

We remark that the signalling dictionary $\mathbf{S}_{i}$ at user
$i$ can be judiciously optimized to suit the preferred choice of
system parameters. In the sequel, we will consider the following construction:
all columns of $\mathbf{S}_{i}$ have equal number of non-zero entries,
set to $\left\lfloor \frac{M}{L}\right\rfloor $, and non-zero entries
are selected uniformly at random from a predefined constellation,
e.g, from the set $\left\{ +1,-1\right\} $. Moreover, every two columns
in $\mathbf{S}_{i}$ have disjoint support. This way, as the transmitted
codeword $\mathbf{x}_{i}$ is formed as a weighted sum of exactly
$l$ columns in $\mathbf{S}_{i}$, the transmitted codeword will have
exactly $l\cdot\left\lfloor \frac{M}{L}\right\rfloor $ non-zero entries,
implying that every user $i$ will have exactly $l\cdot\left\lfloor \frac{M}{L}\right\rfloor $
on-slots and will use its $\tilde{M}=M-l\cdot\left\lfloor \frac{M}{L}\right\rfloor $
off-slots to listen to the incoming signals of other users. Another
way to construct a signalling dictionary would be to apply a regular
Gallager construction, which was originally developed for LDPC codes
(cf., e.g., Ch. VI of \cite{MacKay_book} and references therein). 

Figure \ref{fig:encoder-flow} depicts a CCSM encoder at user $i$.
The encoding three-step procedure is summarized below%
\footnote{Throughout the paper, for $a,b\in\mathbb{N}$, $a\leq b$, $a:b$
denotes the set $\left\{ a,a+1,\ldots,b\right\} $, and for a vector
$\mathbf{x}$, and set of indices $A$, $\mathbf{x}_{A}:=(\mathbf{x}_{a})_{a\in A}$.%
}:
\begin{enumerate}
\item User $i$ encodes $\mathbf{b}_{i}:=\phi_{\mathcal{C}}(\omega_{i,1:k})$
using a CWC code.
\item Further $lq$ bits are encoded on non-zero entries in $\mathbf{b}_{i}$,
i.e., $\mathbf{c}_{i}:=\phi_{w}(\omega_{i,k+1:k+lq},\mathbf{b}_{i})=\phi_{w}(\omega_{i,k+1:k+lq},\phi_{\mathcal{C}}(\omega_{i,1:k}))$.
This is based on a bijective map that assigns a different complex
number to each binary sequence of length $q$, which can be thought
of as a QAM modulation with $2^{q}$ constellation points.
\item User $i$ transmits $\mathbf{x}_{i}=\mathbf{S}_{i}\mathbf{c}_{i}$,
where the matrix-vector multiplication $\mathbf{S}_{i}\mathbf{c}_{i}$
is performed over $\mathbb{C}$. 
\end{enumerate}

\subsection{CCSM Decoder}

\begin{figure}
\noindent \begin{centering}
\includegraphics[bb=40bp 0bp 720bp 470bp,clip,width=3.2in]{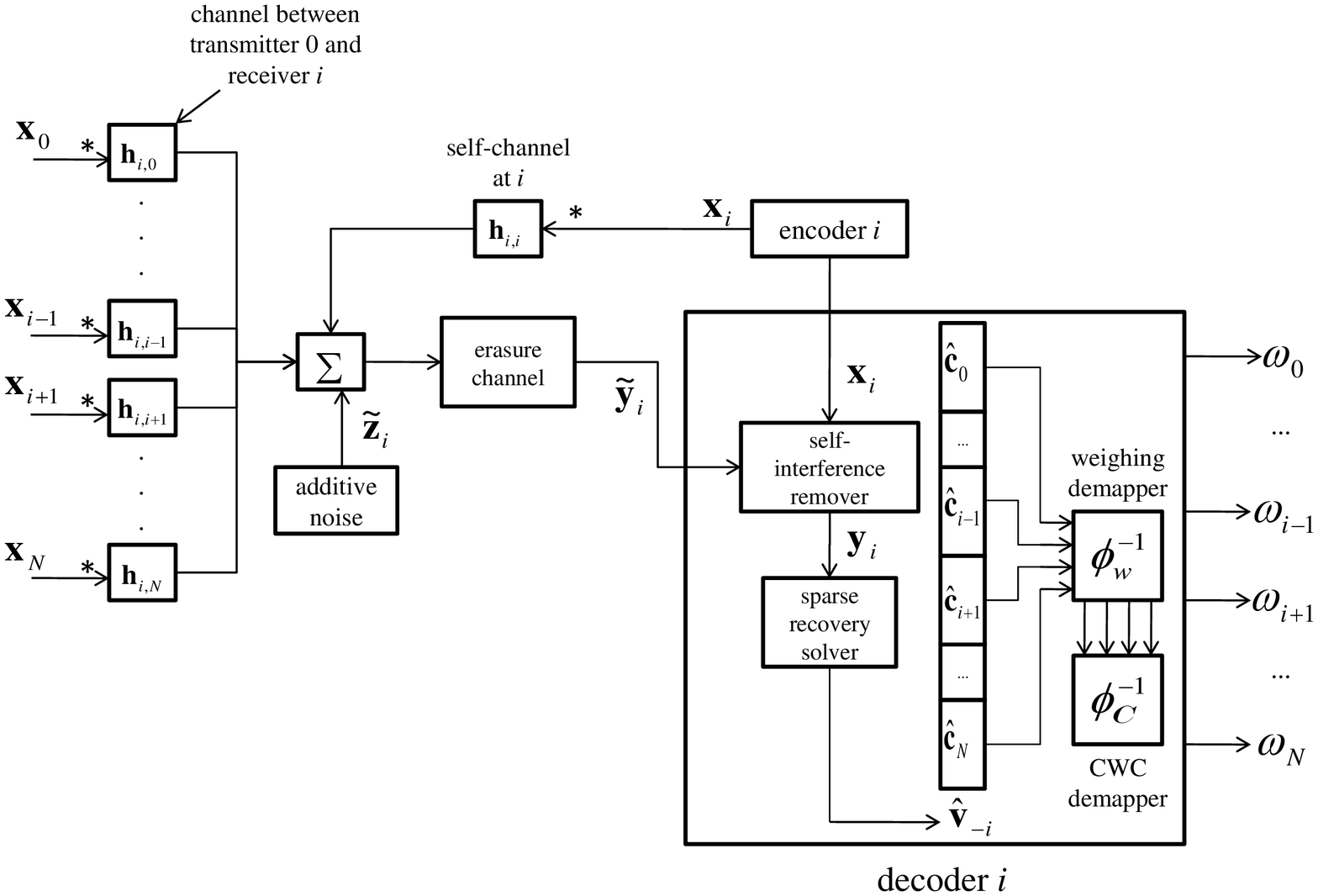}
\par\end{centering}

\caption{CCSM decoder for at $i$-th user.\label{fig:decoder-flow}}

\end{figure}

Figure \ref{fig:decoder-flow} depicts a CCSM decoder at user $i$.
The CCSM decoder receives a superposition of all signals from all
intended transmitters, i.e., users $j\neq i$. As aforementioned,
the receiver does not receive the signal in on-cycles (when it transmits),
which is represented by the erasure channel. Upon removing the self
interference components, the CCSM decoder employs a sparse recovery
solver.

Specifically, the recovery at node $i$ proceeds as follows:
\begin{enumerate}
\item Define an erasure pattern vector as $\mathbf{e}_{i}=\sim\mathbf{1}\left(\mathbf{x}_{i}\right)$,
where $\mathbf{1}\left(\upsilon\right)=0$ if $\upsilon=0$ and $\mathbf{1}\left(\upsilon\right)=1$
otherwise. Define an erasure matrix $\mathbf{E}_{i}$, produced from
$\mathbf{I}_{M,M}$ identity matrix, by removing rows where corresponding
$\mathbf{e}_{i}$ has zero entry. Denote the number of rows in $\mathbf{E}_{i}$
by $\tilde{M}$.
\item User $i$ using off-duty cycles receives:

\begin{equation}
\mathbf{\tilde{y}}_{i}=\mathbf{E}_{i}\left(\sum_{j=0,j\not=i}^{N}\mathbf{h}_{i,j}*\mathbf{S}_{j}\mathbf{c}_{j}+\mathbf{\tilde{z}}_{i}\right)+\mathbf{E}_{i}\left(\mathbf{h}_{i,i}*\mathbf{S}_{i}\mathbf{c}_{i}\right),\label{eq:recsig}\end{equation}

where $"*"$ symbol denotes convolution truncated to $M$ time slots
and $\mathbf{\tilde{z}}_{i}$ represents the additive Gaussian noise
over $M$ time slots. 

\item Since each user switches into reception mode in between transmitting
short bursts, there would be echoes of its own transmitted signal
in the received signal (self interference). However, all users know
their own transmitted signal and can therefore subtract the term $\mathbf{E}_{i}\left(\mathbf{h}_{i,i}*\mathbf{S}_{i}\mathbf{c}_{i}\right)$
in eq. (\ref{eq:recsig}) as long as they know the {}``self channel''.

\selectlanguage{british}%
\begin{equation}
\mathbf{\mathbf{y}}_{i}=\mathbf{\tilde{y}}_{i}-\mathbf{E}_{i}\left(\mathbf{h}_{i,i}*\mathbf{S}_{i}\mathbf{c}_{i}\right)=\mathbf{A}_{-i}\mathbf{v}_{-i}+\mathbf{z}_{i},\label{eq:erasusig}\end{equation}

\selectlanguage{english}%
where $\mathbf{z}_{i}=\mathbf{E}_{i}\mathbf{\tilde{z}}_{i}$, $\mathbf{v}_{-i}$
is the $NL$-column vector formed by concatenating vertically $\mathbf{c}_{0}$,
$\mathbf{c}_{1}$,..., $\mathbf{c}_{i-1}$,$\mathbf{c}_{i+1}$,...$\mathbf{c}_{N},$
i.e., $\mathbf{v}_{-i}=\left[\mathbf{c}_{0}^{\top}|\mathbf{c}_{1}^{\top}|\ldots|\mathbf{c}_{i-1}^{\top}|\mathbf{c}_{i+1}^{\top}|\ldots|\mathbf{c}_{N}^{\top}\right]^{\top}$and
$\mathbf{A}_{-i}$ is an $\tilde{M}\times NL$ matrix, given by:

\begin{eqnarray}
\mathbf{A}_{-i} & = & \mathbf{E}_{i}\biggl[\mathbf{h}_{i,0}*\mathbf{S}_{0}|\mathbf{h}_{i,1}*\mathbf{S}_{1}|\cdots\label{eq: sys_matrix}\\
 &  & |\mathbf{h}_{i,i-1}*\mathbf{S}_{i-1}|\mathbf{h}_{i,i+1}*\mathbf{S}_{i+1}|\cdots|\mathbf{h}_{i,N}*\mathbf{S}_{N}\biggr].\nonumber \end{eqnarray}

Note that the matrix $\mathbf{A}_{-i}$ can be calculated offline,
as it depends only on the channel impulse responses and the signaling
dictionaries. Therefore, it needs to be updated only when the channel
impulse response changes. 

\item User $i$ needs to solve the following problem to detect the desired
signal:

\begin{eqnarray}
\mathbf{\hat{v}}_{-i} & = & \arg\min_{\mathbf{v}_{-i}}\left\Vert \mathbf{\mathbf{y}}_{i}-\mathbf{A}_{-i}\mathbf{v}_{-i}\right\Vert _{2}\nonumber \\
\textrm{s.t.}\;\left\Vert \mathbf{c}_{j}\right\Vert _{0} & = & l,\;\textrm{for all}\; j\neq i,\label{eq: L0_pe}\end{eqnarray}

This is a non-convex optimisation problem. However, we note that exactly
$Nl$ out of $NL$ entries in $\mathbf{v}_{-i}$ are non zero, hence
its sparsity level is by the initial assumption $\frac{l}{L}\ll1$.
This set-up is found in compressive sensing (CS) problems, and thus
one can apply a range of efficient sparse recovery solvers available
in the literature to find an approximate solution to eq. (\ref{eq: L0_pe}),
which we discuss in the next Section.

\item Finally, user $i$ decodes the messages for all $j\neq i$:

\begin{eqnarray*}
(\hat{\omega}_{j,k+1:k+lq},\mathbf{\hat{b}}_{j}) & = & \phi_{w}^{-1}(\mathbf{\hat{c}}_{j}),\\
\hat{\omega}_{j,1:k} & = & \phi_{\mathcal{C}}^{-1}(\mathbf{\hat{b}}_{j}).\end{eqnarray*}

\end{enumerate}

\section{Sparse Recovery for CCSM}

We recall that each user $i$ is required to solve the sparse recovery
problem \eqref{eq: L0_pe} in order to correctly detect the transmitted
messages. This is a non-convex and intractable optimization problem.
However, in the spirit of the compressed sensing framework, one can
apply a convex relaxation, by replacing the $L_{0}$ norm with the
$L_{1}$ norm:

\begin{eqnarray}
\mathbf{\hat{v}}_{-i} & = & \arg\min_{\mathbf{v}_{-i}}\left\Vert \mathbf{\mathbf{y}}_{i}-\mathbf{A}_{-i}\mathbf{v}_{-i}\right\Vert _{2}\nonumber \\
\textrm{s.t.}\;\left\Vert \mathbf{c}_{j}\right\Vert _{1} & = & l,\;\textrm{for all}\; j\neq i.\label{eq: L1_pe}\end{eqnarray}
We will refer to the convex relaxation in \ref{eq: L1_pe} as Group
Basis Pursuit (GBP). Furthermore, one can employ an even simpler form
of the convex relaxation, i.e., a standard embodiment of the LASSO/Basis
Pursuit (BP): 

\begin{eqnarray}
\mathbf{\hat{v}}_{-i} & = & \arg\min_{\mathbf{v}_{-i}}\left\Vert \mathbf{\mathbf{y}}_{i}-\mathbf{A}_{-i}\mathbf{v}_{-i}\right\Vert _{2}\nonumber \\
\textrm{s.t.}\;\left\Vert \mathbf{\mathbf{v}}_{-i}\right\Vert _{1} & \leq & lN,\label{eq: BP_pe}\end{eqnarray}
where the group structure of non-zero entries in $\mathbf{v}_{-i}$
is omitted, but can be enforced after solving \eqref{eq: BP_pe}.

Another method to solve our original problem \eqref{eq: L0_pe}, is
to employ a greedy iterative sparse recovery algorithm. A number of
such algorithms have appeared in the literature including Compressive
Sampling Matching Pursuit (CoSaMP) \cite{cosamp} and Subspace Pursuit
(SP) \cite{SP}. These algorithms can be enhanced to take into account
the additional group structure of the unknown vector, which is imposed
by our system set-up. Namely, in addition to the unknown vector $\mathbf{\mathbf{v}}_{-i}$
having $lN$ non-zero entries, each of its $N$ subvectors $\mathbf{c}_{j}$
of length $L$, has exactly $l$ non-zero entries. In Algorithm \ref{alg: GSP},
we present the modification of Subspace Pursuit, which we name Group
Subspace Pursuit (GSP). For simplicity and without loss of generality,
we present the GSP as applied to the sparse recovery problem at user
$i=0$. The GSP is a low complexity method, which has computational
complexity of Least Square estimator of size $lN$, and is vastly
more computationally efficient than convex optimisation based methods,
including Group Basis Pursuit (GBP) and Basis Pursuit (BP).

\begin{algorithm}
\footnotesize
\begin{itemize}
\item \textbf{Input}. A waveform $\mathbf{y}_{0}\in\mathbb{C}^{\tilde{M}}$
at user $0$, received during the off-duty cycles, with the self-interference
component removed, CIR/Signaling matrix $\mathbf{A}_{-0}\in\mathbb{C}^{\tilde{M}\times NL}$,
CCSM parameters $L$ and $l$.
\item \textbf{Output}. Vector $\mathbf{\hat{v}}_{-0}$ consisting of $N$
subvectors $\mathbf{c}_{j}$ of length $L$, each having exactly $l$
non-zero entries.\end{itemize}
\begin{enumerate}
\item \textbf{Initialize}. Set $\mathbf{r}_{0}=\mathbf{y}_{0}$, $t=1$,
$\mathcal{T}_{0}=\textrm{Ø}$. 
\item \textbf{Identify}. For each $j=1,\ldots,N$, set $\mathcal{U}_{j}$
to the $l$ indices largest in magnitude in the $j$-th $L$-sub-vector
of $\mathbf{A}_{-0}^{*}\mathbf{r}_{t-1}$, i.e.,

\begin{eqnarray*}
\mathcal{U}_{j} & \in & \arg\max_{\mathcal{W}}\Biggl\{\sum_{w\in\mathcal{W}}\left|\left\langle \mathbf{r}_{t-1},\mathbf{a}_{-0,w}\right\rangle \right|:\\
 &  & \mathcal{W}\subset[(j-1)L+1\,:\, jL],\,|\mathcal{W}|=l\Biggr\}.\end{eqnarray*}

\item \textbf{Merge}. Put the old and new columns into one set: $\mathcal{U}=\mathcal{T}_{t-1}\cup\left(\bigcup_{j=1}^{N}\mathcal{U}_{j}\right)$.
\item \textbf{Estimate}. Solve the least-squares problem on the chosen column-set:\begin{eqnarray*}
\mathbf{v}'_{\mathcal{U}} & = & \arg\min_{\mathbf{v}}\left\Vert \mathbf{A}_{-0,\mathcal{U}}\mathbf{\mathbf{v}}-\mathbf{r}_{t-1}\right\Vert _{2}\\
\mathbf{v}'_{[1:N]\backslash\mathcal{U}} & = & \mathbf{0}\end{eqnarray*}

\item \textbf{Prune}. Retain the $l$ coefficients largest in magnitude
in each $L$-sub-vector of $\mathbf{v}'$, i.e.,

\begin{eqnarray*}
\mathcal{U}_{j} & \in & \arg\max_{\mathcal{W}}\Biggl\{\sum_{w\in\mathcal{W}}\left|\mathbf{v}'_{w}\right|:\\
 &  & \mathcal{W}\subset[(j-1)L+1\,:\, jL],\,|\mathcal{W}|=l\Biggr\}.\end{eqnarray*}

to obtain the support estimate $\mathcal{T}_{t}=\bigcup_{j=1}^{N}\mathcal{U}_{j}.$

\item \textbf{Iterate}. Find the $t$-th estimate and update the residual:\begin{eqnarray*}
\mathbf{v}{}_{t,\mathcal{\mathcal{T}}_{t}} & = & \arg\min_{\mathbf{v}}\left\Vert \mathbf{A}_{-0,\mathcal{\mathcal{T}}_{t}}\mathbf{\mathbf{v}}-\mathbf{r}_{t-1}\right\Vert _{2}\\
\mathbf{v}_{t,[1:N]\backslash\mathcal{\mathcal{T}}_{t}} & = & \mathbf{0}\\
\mathbf{r}_{t} & = & \mathbf{y}_{0}-\mathbf{A}_{-0}\mathbf{v}_{t}\end{eqnarray*}
 Set $t\leftarrow t+1$ and repeat (2)-(6) until stopping criterion
holds.
\item \textbf{Output}. Return $\mathbf{\hat{v}}_{-0}=\mathbf{v}_{t}$.
\end{enumerate}
\caption{\label{alg: GSP}Group Subspace Pursuit (GSP).}

\end{algorithm}

\begin{figure}
\noindent \begin{centering}
\includegraphics[scale=0.63]{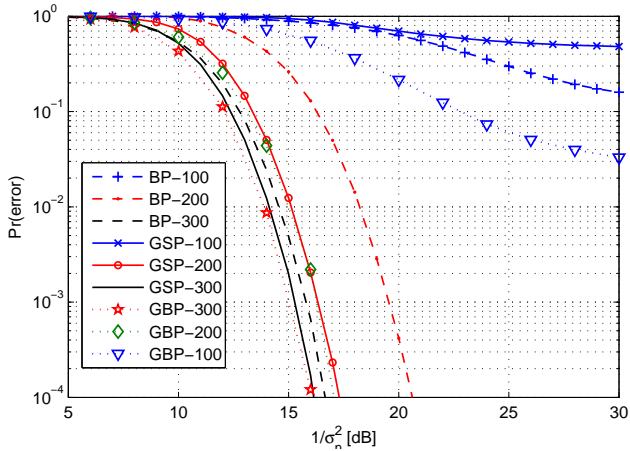}
\par\end{centering}

\caption{Performance comparison of Basis Pursuit/Lasso (BP), Group Basis Pursuit
(GBP) and Group Subspace Pursuit (GSP).\label{fig:GSPvBP}}

\end{figure}

Figure \ref{fig:GSPvBP} depicts performance of the three sparse solvers
for group CS set-up. In this study there are $(N+1)=10$ groups, and
in each group $l=4$ out of $L=32$ elements are non zero. This investigation
was performed for three under sampling ratios for each of those reconstruction
methods. For example, BP-100 signifies the Basis Pursuit solver on
a Complex Gaussian dense measurement matrix with size $100\times320$
(i.e. 31\% under sampling ratio). The non-zero elements in the unknown
vector are drawn from a QPSK modulation set. The error event is defined
as any symbol error in the group. For low under sampling ratios, Group
Basis Pursuit performs best. However, for moderate and larger values,
our Group Subspace Pursuit is almost the same. Therefore, given its
low complexity, we apply GSP to analyse the CCSM performance in the
sequel.

\section{Numerical results}

\begin{figure}
\noindent \begin{centering}
\includegraphics[scale=0.62]{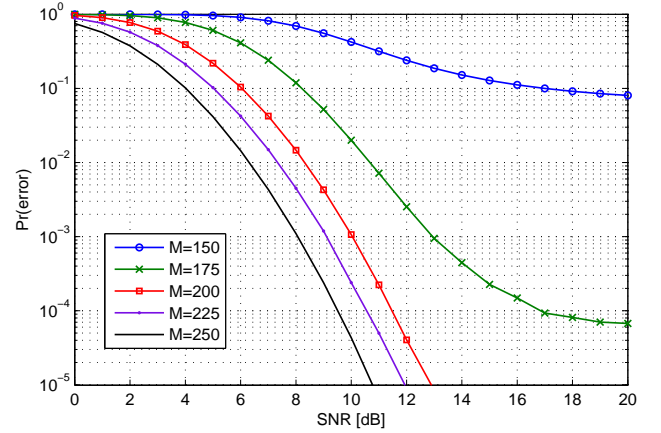}
\par\end{centering}

\caption{Performance of the proposed method in terms of message erorr rate:
In this case there are $(N+1)=5$ users simultaneously broadcasting
messages using CCSM with $L=64,\, l=12$.\label{fig:MER_K5_L64_l12_flat}}

\end{figure}

In this section we report numerical results of the proposed method
and quantitative comparison with the state-of-the-art. We consider
a multi-user wireless network with $N+1$ nodes, where all users are
within radio range of each other. All users attempt to broadcast a
message to all other nodes. We assume a very dispersive channel, modelled
by an FIR filter with 32 taps, with exponentially decaying profile.
Moreover, we assume that each pair of nodes has an independent channel.
We set $L=64$, $l=12$, and use QPSK signalling ($q=2$), i.e., each
message contains $\left\lfloor \log_{2}{L \choose l}\right\rfloor +lq=65$
bits. Figure \ref{fig:MER_K5_L64_l12_flat} depicts the performance
of the proposed method for 5 users in terms of message error rate
(\emph{MER}) as a function of signal-to-noise ratio, for various values
of the number $M$ of available symbol intervals. The \emph{MER} is
an empirical probability estimate of a failure occurring in the message
delivery. We remark that the values of \emph{MER} could be further
decreased by the use of outer coding.

\begin{figure}
\noindent \begin{centering}
\includegraphics[scale=0.47]{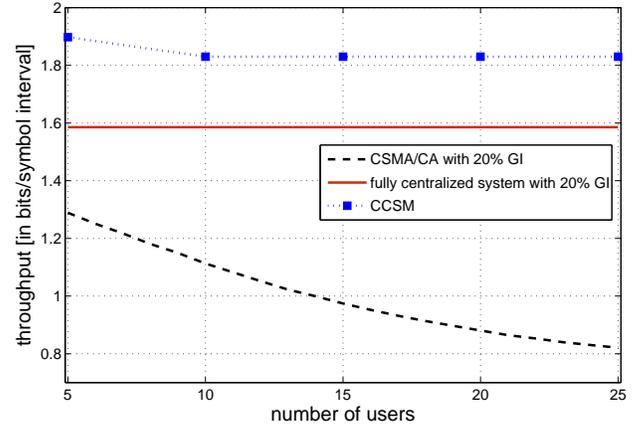}
\par\end{centering}

\caption{\label{fig:Throughput}Throughput comparison of the CCSM and the idealised
versions of CSMA/CA and fully centralized TDMA with guard intervals. }

\end{figure}

To further assess the performance of the CCSM method, we compare its
achieved throughput to the throughput estimates of what would be the
best hypothetical solutions, constructed using the state-of-the-art
in an idealised scenario. As before, we assume that the transmission
occurrs over a time dispersive channel, modelled by an FIR filter
with 32 taps, but, in order to make a fair comparison to MAC protocols
below, without any additive noise. Achieved throughput of CCSM in
bits per symbol interval, given by $(N+1)\cdot65/M_{\min}$, where
$M_{\min}$ is the minimum number of symbol intervals at which no
message errors occurred in at least 100,000 simulation trials, is
plotted in Fig. \ref{fig:Throughput}. We note that the throughput
performance of the CCSM is insensitive to the number of users in the
network.

First hypothetical system we consider exploits a central controlling
mechanism that closely coordinates transmissions between all users,
using a TDMA channel access. To avoid interference the total transmission
time would be divided equally into $N+1$ non overlapping slots. Each
user would broadcast its message to all other users in its designated
slot, and receive messages from all other users in remaining $N$
slots. To cope with the dispersive channel nature, such system would
need to use FDE/OFDM. A typical FDE/OFDM system requires a guard interval
(cyclic prefix) of about 20\% slot duration. However, in reality,
additional guard intervals would be needed, and close coordination
between nodes implies additional overheads. When compared even to
this idealised and highly impractical system, our method offers a
better throughput, as each message transmission requires $\left\lceil \frac{65/2}{0.8}\right\rceil =41$
symbol intervals, which results in the throughput of $65/41=1.58$
bits per symbol interval regardless of the number of users in the
network.

However, in most cases, such a central controlling mechanism would
be unavailable, and the second, more realistic, benchmarking system
we consider is based instead on distributed coordination function
(DCF) and CSMA/CA \cite{Bianchi}, more specifically on DCF as used
in IEEE 802.11b MAC in broadcasting mode. Such system relies on the
randomised deferment of transmissions in order to avoid collisions
on a shared wireless medium. Since we assume that all users are within
radio range of each other, there is no inefficiency resulting from
hidden/exposed terminals, thus we employ only the basic access mechanism
of CSMA/CA protocol. In addition, we assume that each message transmission
contains a guard interval of about 20\% slot duration to cope with
the dispersive channel nature, so that each message transmission requires
41 symbol intervals as above. The minimum and maximum contention windows
of CSMA/CA are assumed to consist of 16 and 1024 symbol intervals,
respectively. We consider an idealised version of the protocol where
no symbol intervals are wasted on distributed or short interframe
space (DIFS/SIFS), propagation delay, physical or MAC message headers
and ACK responses. Moreover, the transmission queue of each user consists
of a single message. Thus, any inefficiency of the scheme is a result
either of the idle contention intervals or collisions. The simulated
average throughput of such scheme is presented in Fig. \ref{fig:Throughput}.
We note that CCSM significantly outperforms even such idealised CSMA/CA
scheme offering, e.g., twice the throughput of idealised CSMA/CA in
the case of 20 users.

\section{Conclusions}

In this paper we have introduced a novel modulation and multiplexing
method for ad-hoc wireless networks. The CCSM method offers a range
of benefits: same time/frequency duplex, minimal MAC, inherent robustness
in time dispersive channels. The CCSM is also applicable to optical
communications (both guided and free space), where it could offer
better performance/flexibility than combinatorial PPM. We have demonstrated
significant data throughput improvements against the state-of-the
art. However, the presented performance gains of CCSM are conservative,
since we have opted for a low complexity detection method. Further
performance gains can be achieved by employing sparse recovery methods
which would capitalise on the discrete nature of the unknown signal
vector.

\end{document}